\documentclass[a4paper,10pt,twoside]{cpc-hepnp}
\usepackage{CJK,upgreek,fancyhdr}
\usepackage{multicol}
\usepackage{graphicx}
\usepackage{booktabs}
\usepackage{amssymb,bm,mathrsfs,bbm,amscd}
\usepackage[tbtags]{amsmath}
\usepackage{lastpage}

\begin{document}
\begin{CJK*}{GBK}{song}

\fancyhead[c]{\small Chinese Physics C~~~Vol. xx, No. x (201x) xxxxxx}
\fancyfoot[C]{\small 010201-\thepage}

\footnotetext[0]{Received 31 June 2015}

\title{Anisotropic compact stars in Karmarkar spacetime}

\author{%
      Ksh. Newton Singh$^{1;1)}$\email{ntnphy@gmail.com}%
\quad Neeraj Pant$^{2;2)}$\email{neeraj.pant@yahoo.com}%
\quad M. Govender$^{3;3}$\email{Govenderm43@ukzn.ac.za}
}
\maketitle

\address{%
$^1$ Department of Physics, National Defence Academy, Khadakwasla, Pune-411023, India.\\
$^2$ Department of Mathematics, National Defence Academy, Khadakwasla, Pune, 411023, India.\\
$^3$ Department of Mathematics, Faculty of Applied Sciences, Durban University of Technology, Durban, South Africa. 
}

\begin{abstract}
We present a new class of solutions to the Einstein field equations for an anisotropic matter distribution in which the interior space-time obeys the Karmarkar condition. The necessary and sufficient condition required for a spherically symmetric space-time to be of class one reduces the gravitational behavior of the model to a single metric function. By assuming a physically viable form for the $g_{rr}$ metric potential we obtain an exact solution of the Einstein field equations which is free from any singularities and satisfies all the physical criteria. We utilize this solution to predict the masses and radii of well-known compact objects such as Cen X-3, PSR J0348+0432, PSR B0943+10 and XTE J1739-285.
\end{abstract}



\footnotetext[0]{\hspace*{-3mm}\raisebox{0.3ex}{$\scriptstyle\copyright$}2013
Chinese Physical Society and the Institute of High Energy Physics
of the Chinese Academy of Sciences and the Institute
of Modern Physics of the Chinese Academy of Sciences and IOP Publishing Ltd}%

\begin{multicols}{2}

\section{Introduction}
The century-old search for exact solutions of the Einstein field equations began with Karl Schwarzschild obtaining vacuum solution describing the exterior of a spherically symmetric matter distribution \cite{karl1}. A natural line of pursuit would be to find an interior solution which matched smoothly to the Schwarzschild exterior solution. This internal solution was obtained by Schwarzschild in which assumed that the internal matter content of a spherical mass distribution was characterized by uniform density \cite{karl2}. Observations of stars and the understanding of particle physics within dense cores necessitated the search for more realistic solutions of the field equations. The inclusion of pressure anisotropy, charge, bulk viscosity, an equation of state, multilayered fluids and the departure from spherical symmetry has led to the discovery of hundreds of exact solutions describing relativistic stars in the static limit \cite{bowers,beken,sunil1,tik1,escu}. With the discovery of the Vaidya solution, it became necessary to model the gravitational collapse of radiating stars \cite{vaidya}. Since the star is dissipating energy in the form of a radial heat flux, the pressure at the boundary of the star is proportional to the outgoing heat flux as opposed to vanishing surface pressure in the non-dissipative case \cite{santos}. Nevertheless, static solutions also play a pivotal role in dissipative gravitational collapse of stars as they can represent an initial static configuration or a final static configuration \cite{bonnor,sharma1,sharma2}.

It is interesting to note that by relaxing the condition of a perfect fluid and allowing for pressure anisotropy and charge within the interior of the stellar distribution gives rise to observable and measurable properties of the star. Pressure anisotropy leads to arbitrarily large surface red-shifts   \cite{bhar1,bhar2,singh1} while the inclusion of charge results in the modification of the Buchdahl limit (\cite{andy}). The linear equation of state $p=\alpha \rho$ has been generalized from observations in theoretical particle physics. There has been a wide spectrum of exact solutions of the field equations incorporating the so-called MIT bag model in which the equation of state is of the form $p=\alpha \rho-B$ with $B$ being the bag constant \cite{sunil2,sunil3,thiruk}. These solutions successfully predicted the observed masses and radii of compact objects with densities of the order of $10^{14} g~cm^{-3}$. With an ever growing interest in dark energy and its successful use in cosmological models, astrophysicists have now extended the range of $\alpha$ in $p = \alpha \rho$ to include $-1< \alpha < -1/3$. This regime incorporates the so-called dark stars \cite{bhar10,lobo}. Other exotic forms of matter which have appeared in the literature include the Chaplygin gas, Bose-Einstein condensates and the Hagedorn fluid \cite{far1,far2,bhar3,harko1,harko2}.

The notion of the four fundamental interactions being a manifestation of a single force has always attracted the interest of researchers in both fundamental particles physics and relativity. Higher dimensional theories of gravity have produced rich results in so far as cosmic censorship is concerned  \cite{dad1,pankaj1,dadh}. Recently, there has been a surge in exact models of stars in Einstein-Gauss-Bonnet gravity, braneworld gravity as well as Lovelock gravity \cite{brian1,brian2,sudan1,megann1}. The connection between five-dimensional Kaluza-Klein geometries and electromagnetism has been widely studied. Embedding of four-dimensional space-times into higher dimensions is an invaluable tool in generating both cosmological and astrophysical models. In this paper we utilize the Karmarkar \cite{kar} condition which is a necessary and sufficient condition for a spherically symmetric line element to be of class one to generate exact solutions of the Einstein field equations. In particular, our model incorporates anisotropic pressure and is free from any singularities. Our results support recent observations made by \cite{newton1,newton2,smit} for models in embedding class I.

\section{Interior Space-time}

The interior of the super-dense star is assumed to be described by the line element
\begin{equation}
ds^2 = e^{\nu(r)}dt^2 - e^{\lambda(r)}dr^2 - r^2(d\theta^2 + \sin^2{\theta}d\phi^2) \label{metric}
\end{equation}

The energy-momentum tensor for the stellar anisotropic fluid is
\begin{equation}
T_{ab} = {\mbox diag}\left(\rho, -p_r, -p_t, -p_t\right),\label{2}
\end{equation}

where $\rho$, $p_r$ and $p_t$ are the energy density, radial pressure and tangential pressure, respectively.

The Einstein field equations for the line element (\ref{metric})
are
\begin{eqnarray}\label{g3}
8\pi \rho &=&
\frac{1 - e^{-\lambda}}{r^2} + \frac{\lambda'e^{-\lambda}}{r} \label{g3a} \\ \nonumber \\
8\pi p_r &=&  \frac{\nu' e^{-\lambda}}{r} - \frac{1 - e^{-\lambda}}{r^2} \label{g3b}  \\  \nonumber \\
8\pi p_t &=& \frac{e^{-\lambda}}{4}\left(2\nu'' + {\nu'}^2  - \nu'\lambda' + \frac{2\nu'}{r}-\frac{2\lambda'}{r}\right) \label{g3c}
\end{eqnarray}

where primes represent differentiation with respect to the radial coordinate $r$. In generating the above field equations we have utilized geometrized units where $G$ and $c$ are taken to be unity. Using the Eqs. (\ref{g3b}) and (\ref{g3c}) we obtain the anisotropy parameter 

\begin{eqnarray}
\Delta &=& 8\pi (p_t-p_r) \nonumber \\
& = & e^{-\lambda}\left[{\nu'' \over 2}-{\lambda' \nu' \over 4}+{\nu'^2 \over 4}-{\nu'+\lambda' \over 2r}+{e^\lambda-1 \over r^2}\right]\label{del}
\end{eqnarray}

If the metric given in (\ref{metric}) satisfies the Karmarkar condition\cite{kar} , it can represent a embedding class one spacetime i.e.
\begin{equation}
R_{1414}={R_{1212}R_{3434}+R_{1224}R_{1334} \over R_{2323}}\label{con}
\end{equation}
with $R_{2323}\neq 0$ \cite{pandey}. This condition leads to a differential equation given by

\begin{equation}
{2\nu'' \over \nu'}+\nu'={\lambda' e^\lambda \over e^\lambda-1}\label{dif1}
\end{equation}

On integration we get the relationship between $\nu$ and $\lambda$ as
\begin{equation}
e^{\nu}=\left(A+B\int \sqrt{e^{\lambda}-1}~dr\right)^2\label{nu1}
\end{equation}
where $A$ and $B$ are constants of integration.

By using (\ref{nu1}) we can rewrite (\ref{del}) as
\begin{eqnarray}
\Delta = {\nu' \over 4e^\lambda}\left[{2\over r}-{\lambda' \over e^\lambda-1}\right]~\left[{\nu' e^\nu \over 2rB^2}-1\right] \label{del1}
\end{eqnarray}

\section{Anisotropic stellar solution}

To solve the above equation (\ref{nu1}), we have assumed an entirely new type of $g_{rr}$ metric potential given by
\begin{equation}
e^{\lambda}= {4(1+ar^2)^2 \over (2-ar^2)^2}\label{elam}
\end{equation}

On integrating (\ref{nu1}), we get
\begin{eqnarray}
e^\nu & = & \bigg[A+{B \over \sqrt{a}} \bigg\{\sqrt{12+3ar^2} \nonumber \\
&& - 3\sqrt{2}~\tanh^{-1}\left({\sqrt{4+ar^2} \over \sqrt{6}}\right)\bigg\} \bigg]^2 \label{enu}
\end{eqnarray}

Now using (\ref{enu}) and (\ref{elam}) in (\ref{g3a}), (\ref{g3b}), (\ref{del1}) and (\ref{g3c}), we get
\begin{eqnarray}
8\pi \rho &=& \frac{3 a (a^2 r^4+a r^2+12)}{4(a r^2+1)^3} \label{rho}\\
8\pi p_r &=& {a\sqrt{4+ar^2}\Big(f_3(r)-f_1(r)\sqrt{4+ar^2}\Big) \over 4(1+ar^2)^2\big(f_2(r)-f_4(r)\big)} \label{pr12}\\
\Delta &=& \frac{\sqrt{3} a^2 r^2(a r^2+7) ~\big[f_4(r)-f_2(r)\big]^{-1}}{4 (a r^2+1)^3 \sqrt{a r^2+4}} \times \nonumber\\
&& \bigg(f_6(r)-f_5(r) \sqrt{a r^2+4}\bigg) \label{del2}\\
8\pi p_t &=&  8\pi p_r+\Delta
\end{eqnarray}
\end{multicols}

where
\begin{eqnarray}
f_1(r) & = & 3 A\sqrt{a r^2(a r^2+4)} +\sqrt{3} B r (a r^2+16)\\
f_2(r) & = & A \sqrt{a r^2 (a r^2+4)} +\sqrt{3} B r(a r^2+4)\\
f_3(r) & = & 9 \sqrt{2}~ B r(a r^2+4) \tanh^{-1}\left(\frac{\sqrt{a r^2+4}}{\sqrt{6}}\right)\\
f_4(r) & = & 3 B r \sqrt{2 a r^2+8}~ \tanh^{-1}\left(\frac{\sqrt{a r^2+4}}{\sqrt{6}}\right)\\
f_5(r) & = & \sqrt{3} A \sqrt{a r^2(a r^2+4)} +2 B r(a r^2+7)\\
f_6(r) & = & 3 \sqrt{6} B r(a r^2+4) \tanh^{-1}\left(\frac{\sqrt{a r^2+4}}{\sqrt{6}}\right)
\end{eqnarray}

We can write the density and pressure gradients as

\begin{eqnarray}
{d\rho \over dr} &=& -\frac{3 a^2 r (a^2 r^4+35)}{2(a r^2+1)^4}\\
8\pi {dp_r \over dr} &=& -\frac{a r \sqrt{a r^2}}{2 g_7(x)~ (a r^2+1)^3}\bigg[3 \sqrt{2} B~ \big\{(g_4(x)+g_5(x)\big\} \tanh ^{-1}\left(\frac{\sqrt{a r^2+4}}{\sqrt{6}}\right) \nonumber \\
&& -\sqrt{a r^2+4}~ \big\{g_1(x)+g_2(x)+g_3(x)\big\}-g_6(x)\bigg]\\
8\pi {dp_t \over dr} &=& -\frac{a^3 r^5 \sqrt{a r^2+4} }{4 p_7(x)~ \left(a r^2+1\right)^4 \left(a r^2 \left(a r^2+4\right)\right)^{3/2}}\bigg[-3 \sqrt{2} B~ \big\{p_4(x)+p_5(x)\big\} \nonumber \\
&&  \tanh ^{-1}\left(\frac{\sqrt{a r^2+4}}{\sqrt{6}}\right)+\sqrt{a r^2+4} \big\{p_1(x)+p_2(x)-p_3(x)\big\} +p_6(x)\bigg]\\
\mbox{Where}\\
g_1(x) &=& a^3 r^4 \left[3 A^2 \sqrt{\frac{a r^2 (a r^2+4)}{(a r^2-2)^2}}+3 B^2 r^2 \sqrt{\frac{a r^2(a r^2+4)}{(a r^2-2)^2}}+5 \sqrt{3} A B r\right] \nonumber \\
&& -708 B^2 \sqrt{\frac{a r^2 (a r^2+4)}{(a r^2-2)^2}}\\
g_2(x) &=& a^2 r^2 \bigg[15 A^2 \sqrt{\frac{a r^2 (a r^2+4)}{(a r^2-2)^2}}+81 B^2 r^2 \sqrt{\frac{a r^2 (a r^2+4)}{(a r^2-2)^2}} +67 \sqrt{3} A B r\bigg]\\
g_3(x) &=& 2 a \bigg[-21 A^2 \sqrt{\frac{a r^2 (a r^2+4)}{(a r^2-2)^2}}+90 B^2 r^2 \sqrt{\frac{a r^2 (a r^2+4)}{(a r^2-2)^2}} +103 \sqrt{3} A B r\bigg]\\
g_4(x) &=& 24 a r \left[3 \sqrt{3} B r \sqrt{\frac{a r^2 (a r^2+4)}{(a r^2-2)^2}}+7 A\right] -412 \sqrt{3} B \sqrt{\frac{a r^2 (a r^2+4)}{(a r^2-2)^2}}\\
g_5(x) &=& a^3 r^5 \left[5 \sqrt{3} B r \sqrt{\frac{a r^2 (a r^2+4)}{(a r^2-2)^2}}+6 A\right] +3 a^2 r^3 \left[19 \sqrt{3} B r \sqrt{\frac{a r^2 (a r^2+4)}{(a r^2-2)^2}}+22 A\right]\\
g_6(x) &=& 54 B^2\big\{a^2 r^4+5 a r^2-14\big\} \sqrt{\frac{a r^2 (a r^2+4)^2}{(a r^2-2)^2}}~ \left[\tanh ^{-1}\left(\frac{\sqrt{a r^2+4}}{\sqrt{6}}\right)\right]^2\\
g_7(x) &=& \bigg[A \sqrt{a r^2 (a r^2+4)}+\sqrt{3} B r (a r^2+4) -3 B r \sqrt{2 a r^2+8}~ \tanh ^{-1}\left(\frac{\sqrt{a r^2+4}}{\sqrt{6}}\right)\bigg]^2
\end{eqnarray}
\begin{eqnarray}
p_1(x) &=& 9 a^3 r^4 \left[-4 A^2 \sqrt{\frac{a r^2 (a r^2+4)}{(a r^2-2)^2}}+22 B^2 r^2 \sqrt{\frac{a r^2 (a r^2+4)}{(a r^2-2)^2}}+23 \sqrt{3} A B r\right]+ \nonumber \\
&& a^5 B r^9 \left[6 B r \sqrt{\frac{a r^2 (a r^2+4)}{(a r^2-2)^2}} +\sqrt{3} A\right] +10368 B^2 \sqrt{\frac{a r^2 (a r^2+4)}{(a r^2-2)^2}}\\
p_2(x) &=& 2 a^4 r^6 \left[12 A^2 \sqrt{\frac{a r^2 (a r^2+4)}{(a r^2-2)^2}}+39 B^2 r^2 \sqrt{\frac{a r^2 (a r^2+4)}{(a r^2-2)^2}}+23 \sqrt{3} A B r\right] \nonumber \\
&& -4 a \bigg[-168 A^2 \sqrt{\frac{a r^2 (a r^2+4)}{(a r^2-2)^2}}+795 B^2 r^2 \sqrt{\frac{a r^2 (a r^2+4)}{(a r^2-2)^2}} +776 \sqrt{3} A B r\bigg]\\
p_3(x) &=& 2 a^2 r^2 \bigg[180 A^2 \sqrt{\frac{a r^2 (a r^2+4)}{(a r^2-2)^2}}+879 B^2 r^2 \sqrt{\frac{a r^2 (a r^2+4)}{(a r^2-2)^2}} +337 \sqrt{3} A B r\bigg]\\
p_4(x) &=& \sqrt{3} a^5 B r^{10} \sqrt{\frac{a r^2 (a r^2+4)}{(a r^2-2)^2}}+4 a^4 r^7 \left[11 \sqrt{3} B r \sqrt{\frac{a r^2 (a r^2+4)}{(a r^2-2)^2}}+12 A\right] \nonumber \\
&& +6208 \sqrt{3} B \sqrt{\frac{a r^2 (a r^2+4)}{(a r^2-2)^2}}\\
p_5(x) &=& a^3 r^5 \left[115 \sqrt{3} B r \sqrt{\frac{a r^2 (a r^2+4)}{(a r^2-2)^2}}+216 A\right] -64 a^2 r^3 \left[17 \sqrt{3} B r \sqrt{\frac{a r^2 (a r^2+4)}{(a r^2-2)^2}}+9 A\right] \nonumber \\
&& -4 a r \left[439 \sqrt{3} B r \sqrt{\frac{a r^2 (a r^2+4)}{(a r^2-2)^2}}+672 A\right]\\
p_6(x) &=& 216 B^2 \big\{(2 a^3 r^6-3 a^2 r^4-30 a r^2+56\big\} \sqrt{\frac{a r^2 (a r^2+4)^2}{(a r^2-2)^2}} \left[\tanh ^{-1}\left(\frac{\sqrt{a r^2+4}}{\sqrt{6}}\right)\right]^2\\
p_7(x) &=& \bigg[A \sqrt{\frac{a r^2 (a r^2+4)}{(a r^2-2)^2}} (a r^2-2)+\sqrt{3} B r (a r^2+4)-3 B r \sqrt{2 a r^2+8} \times \nonumber \\
&& \tanh ^{-1}\left(\frac{\sqrt{a r^2+4}}{\sqrt{6}}\right)\bigg]^2
\end{eqnarray}
\ruledown

\begin{multicols}{2}
These density and pressure gradients are represented graphically in Fig. \ref{grd}.

Using the relationship between $e^\lambda$ and mass $m(r)$ i.e.
\begin{equation}
e^{-\lambda}=1-{2m \over r}
\end{equation}
and (\ref{g3a}) we get
\begin{eqnarray}
m(r) &=& 4\pi \int_0^r \rho r^2 dr = \frac{3 a r^3(a r^2+4)}{8(a r^2+1)^2}\label{mf}
\end{eqnarray}

\section{Conditions for physical viability of the solutions}

The following conditions are to be fulfilled by the solution in order to represent a physically viable configuration.

\begin{enumerate}
\item  The solution should be free from physical and geometric singularities, i.e. it should yield finite and positive values of the central pressure, central density and nonzero positive value of  $e^\nu|_{r=0}$ and  $e^\lambda|_{r=0}=1$.

\item  The causality condition should be obeyed i.e. velocity of sound should be less than that of light throughout the model. In addition to the above the velocity of sound should be decreasing towards the surface i.e.${d \over dr}~{dp_r \over d\rho}<0$  or   ${d^2 p_r\over d\rho^2}>0$   and ${d\over dr}{dp_t \over d\rho}<0$  or   ${d^2p_t \over d\rho^2}>0$  for $0\leq r\leq r_b$ i.e. the velocity of sound is increasing with the increase of density and it should be decreasing outwards.

\item  	The adiabatic index,  $\gamma= {\rho+p_r \over p_r}{dp_r \over d\rho}$  for realistic matter should be  $\gamma > 1$.

\item   The red-shift $z$ should be positive, finite and monotonically decreasing in nature with the increase of the radial coordinate.

\item   For a stable anisotropic compact star, $0<|v_t^2-v_r^2|\leq 1$ must be satisfied, \cite{herrera97}.

\item Anisotropy must be zero at the center and increasing outward.

\end{enumerate}
\ruleup

\section{Properties of the solution}

The central values of $p_r,~p_t$, $\rho$ and the Zeldovich's condition can be written as
\end{multicols}
\begin{eqnarray}
8\pi p_{rc} = 8\pi p_{tc} & = & \frac{a }{2}\bigg[-2 \sqrt{a}~ A+4 \sqrt{3} ~B -6 \sqrt{2}~ B \tanh^{-1}(\sqrt{2/3})\bigg] \times \nonumber\\
& & \bigg[36 \sqrt{2}~ B \tanh ^{-1}(\sqrt{2/3}) -32 \sqrt{3} ~B-12 \sqrt{a}~ A\bigg] >0  \label{cons1}\\
8\pi \rho_c &=& 9a > 0 ;~~\forall ~a>0\\
{p_{rc} \over \rho_c} &=& \frac{1 }{18}\bigg[-2 \sqrt{a}~ A+4 \sqrt{3}~ B -6 \sqrt{2}~ B \tanh^{-1}(\sqrt{2/3})\bigg]\times \nonumber\\
& &  \bigg[36 \sqrt{2}~ B \tanh ^{-1}(\sqrt{2/3}) -32 \sqrt{3}~ B-12A \sqrt{a} \bigg] \le 1\label{cons2}
\end{eqnarray}
\ruledown

\begin{multicols}{2}
Now using the two constraints on $A,~B$ and $a$ given in (\ref{cons1}) and (\ref{cons2}), we get a final form as
\begin{eqnarray}
{8\sqrt{3}-9\sqrt{2}~\tanh^{-1}(\sqrt{2/3}) \over 3\sqrt{a}}  <   {A \over B} \nonumber \\
\le  {13\sqrt{3}-18\sqrt{2}~\tanh^{-1}(\sqrt{2/3}) \over 6\sqrt{a}} \label{cons3}
\end{eqnarray}

Now the velocity of sound within the stellar object can be found as
\begin{equation}
v_r^2={dp_r/dr \over d\rho/dr},~~~v_t ^2={dp_t/dr \over d\rho/dr}
\end{equation}

The relativistic adiabatic index and the compression modulus is given by
\begin{equation}
\Gamma_r = {\rho+p_r \over p_r}~{dp_r \over d\rho};~~~\Gamma_t = {\rho+p_t \over p_t}~{dp_t \over d\rho}
\end{equation}
For a static configuration at equilibrium $\Gamma_r$ has to be more than $4/3$.

The generalized Tolman-Oppenheimer-Volkoff (TOV) equation was contributed by \cite{ponce} as
\begin{eqnarray}
-{M_g(\rho+p_r) \over r^2}~e^{(\lambda-\nu)/2}-{dp_r \over dr}+{2(p_t-p_r) \over r}=0 \label{tove}
\end{eqnarray}
provided
\begin{eqnarray}
M_g(r) &=& {1\over 2}~r^2 \nu' e^{(\nu-\lambda)/2}
\end{eqnarray}

The above equation (\ref{tove}) can be written in terms of balanced force equation due to anisotropy ($F_a$), gravity ($F_g$) and hydrostatic ($F_h$) i.e.
\begin{equation}
F_g+F_h+F_a=0 \label{forc}
\end{equation}

Here
\begin{eqnarray}
F_g & = & -{M_g(\rho+p_r) \over r^2}~e^{(\lambda-\nu)/2}\\
F_h &=& -{dp_r \over dr}\\
F_a &=& {2(p_t-p_r) \over r}
\end{eqnarray}
The generalized TOV equation (\ref{forc}) can be represent by a figure showing the forces that are balanced to each  Fig. \ref{tov1}.

\section{Boundary conditions}

Assuming the exterior spacetime is the Schwarzschild solution which has to be match smoothly with the interior solution and is given by
\begin{eqnarray}
ds^2 &=& \left(1-{2M\over r}\right) dt^2-\left(1-{2M\over r}\right)^{-1}dr^2 \nonumber \\
&& -r^2(d\theta^2+\sin^2 \theta d\phi^2) \label{ext}
\end{eqnarray}

By matching the interior solution (\ref{metric}) and exterior solution (\ref{ext}) at the boundary $r=r_b$ we get
\begin{eqnarray}
e^{\nu_b} &=& 1-{2M \over r_b} \nonumber \\
& = & \bigg[A+{B \over \sqrt{a}} \bigg\{\sqrt{12+3ar_b^2}- 3\sqrt{2} \times  \nonumber \\
&& \tanh^{-1}\left(\sqrt{{4+ar_b^2 \over 6}}\right)\bigg\} \bigg]^2\label{bou1}\\
e^{-\lambda_b} &=& 1-{2M \over r_b} = {\big(2-ar_b^2\big)^2 \over 4\big(1+ar_b^2\big)^2} \label{bou2}\\
p_r(r_b) &=& 0 \label{bou3}
\end{eqnarray}

\begin{center}
\includegraphics[width=5cm]{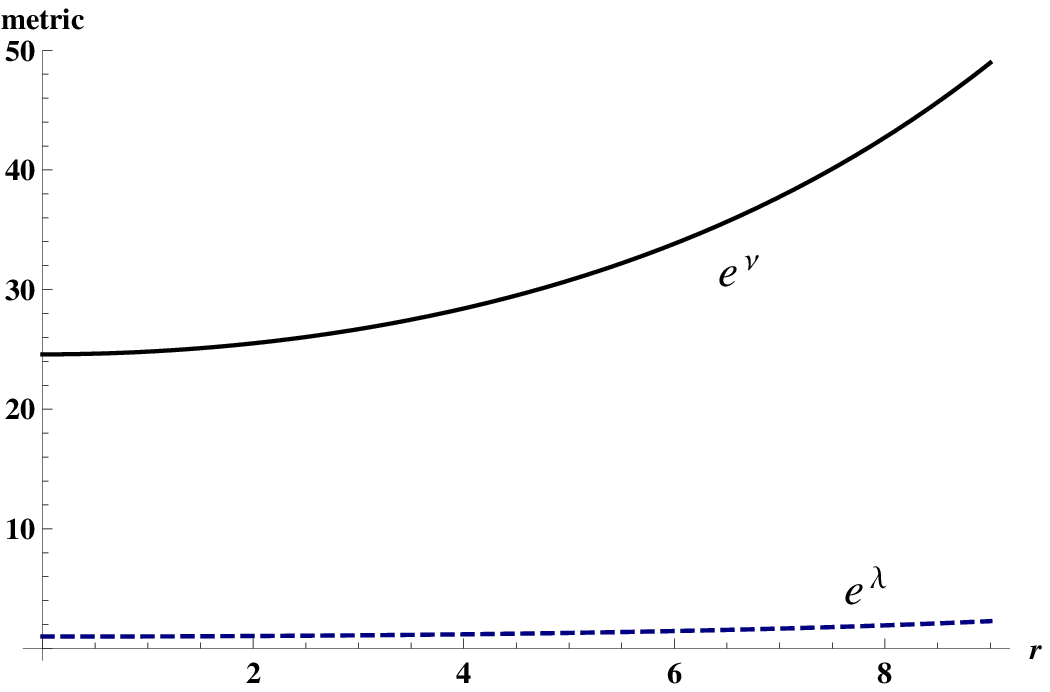}
\figcaption{\label{met}   Variation of metric potential with radius.}
\end{center}

\begin{center}
\includegraphics[width=5cm]{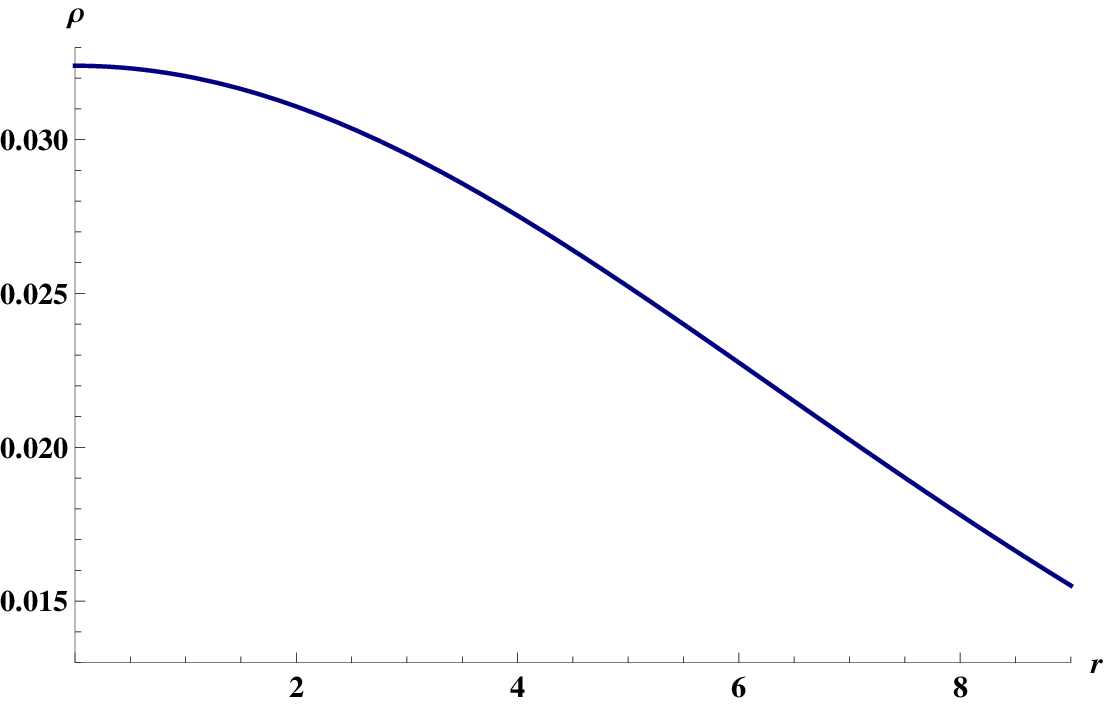}
\figcaption{\label{rho1}   Variation of density with radius.}
\end{center}

\begin{center}
\includegraphics[width=5cm]{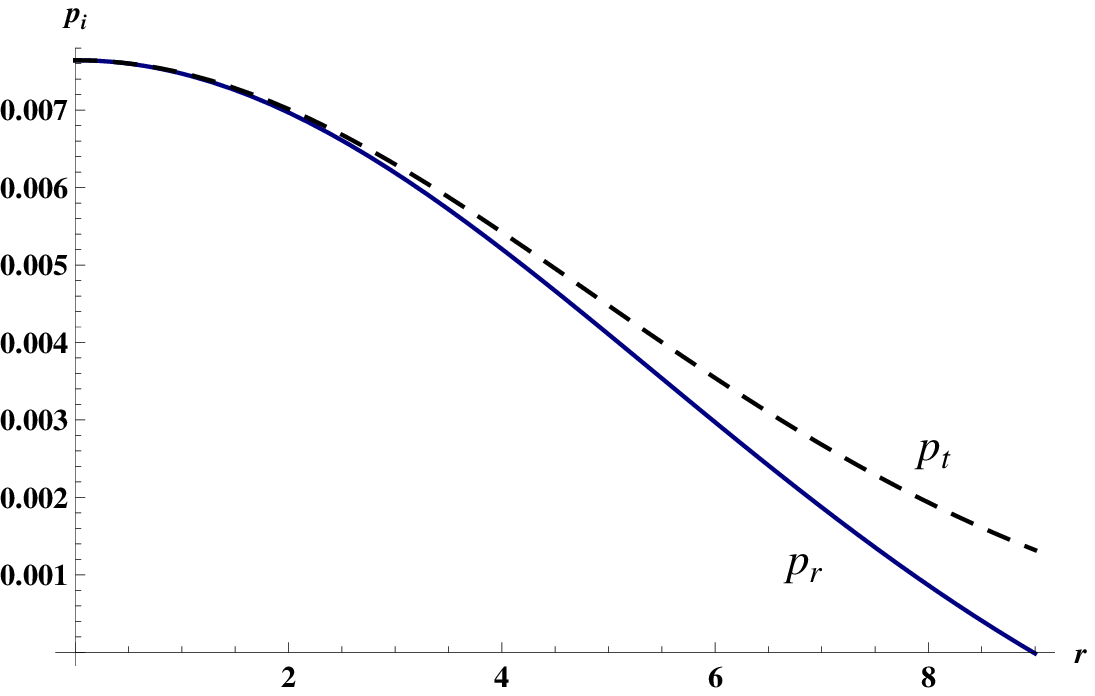}
\figcaption{\label{pres}   Variation of pressures with radius.}
\end{center}

\begin{center}
\includegraphics[width=5cm]{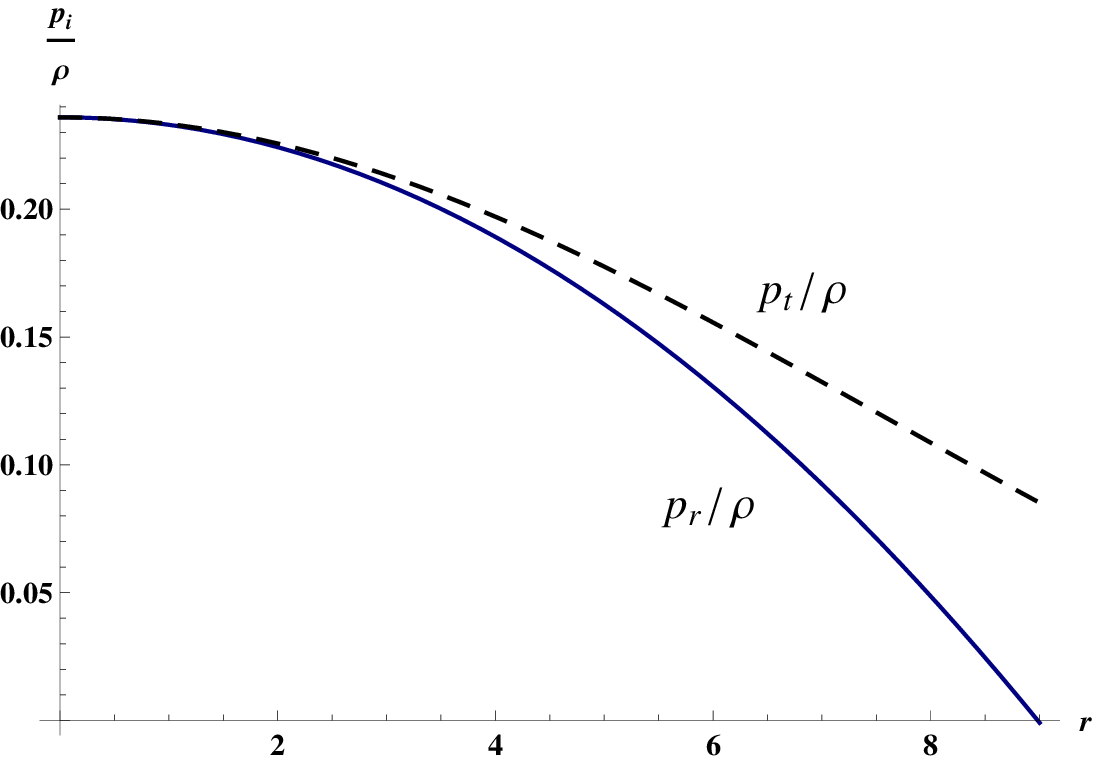}
\figcaption{\label{perho}   Variation of pressure to density ratios with radius.}
\end{center}

\begin{center}
\includegraphics[width=5cm]{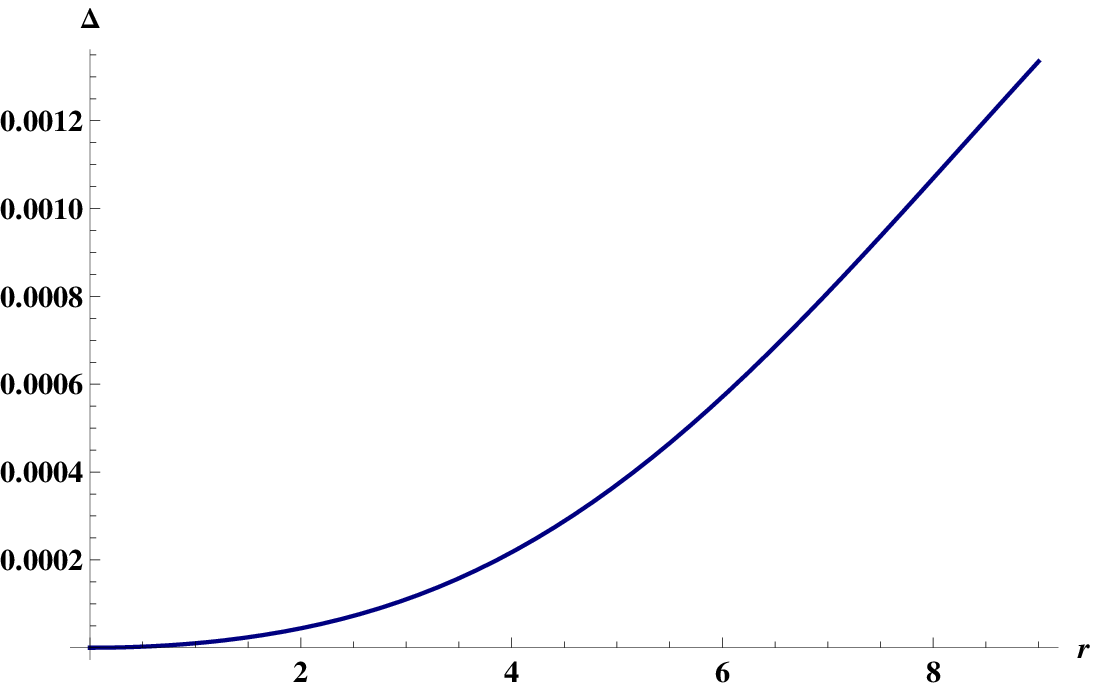}
\figcaption{\label{ani}   Variation of anisotropy with radius.}
\end{center}

\begin{center}
\includegraphics[width=5cm]{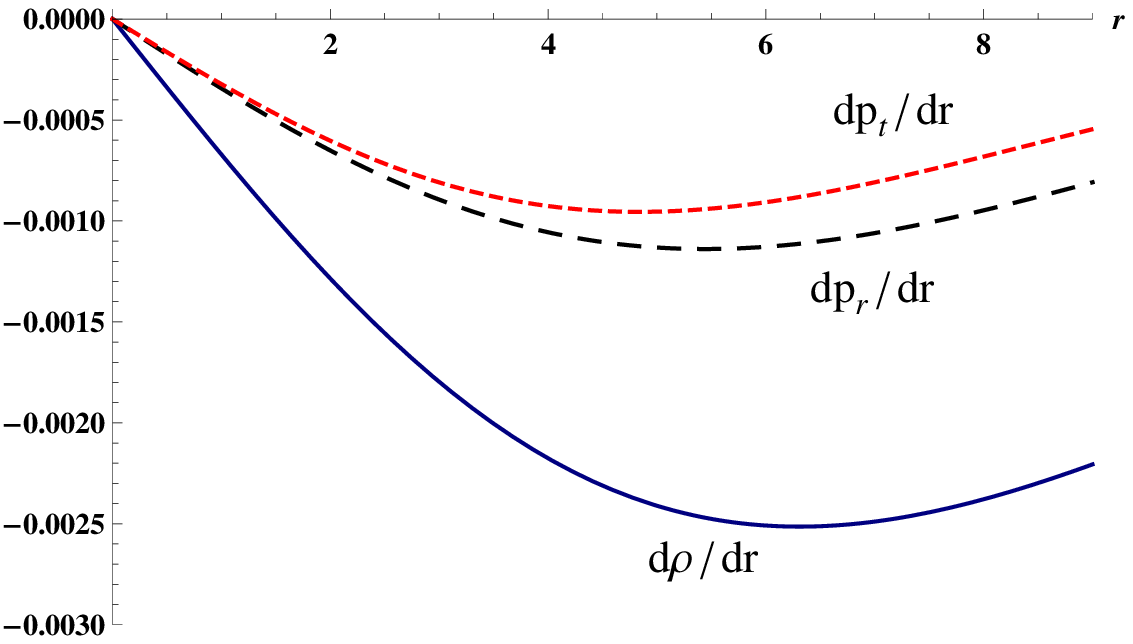}
\figcaption{\label{grd}   Variation of pressure and density gradients with radius.}
\end{center}

\begin{center}
\includegraphics[width=5cm]{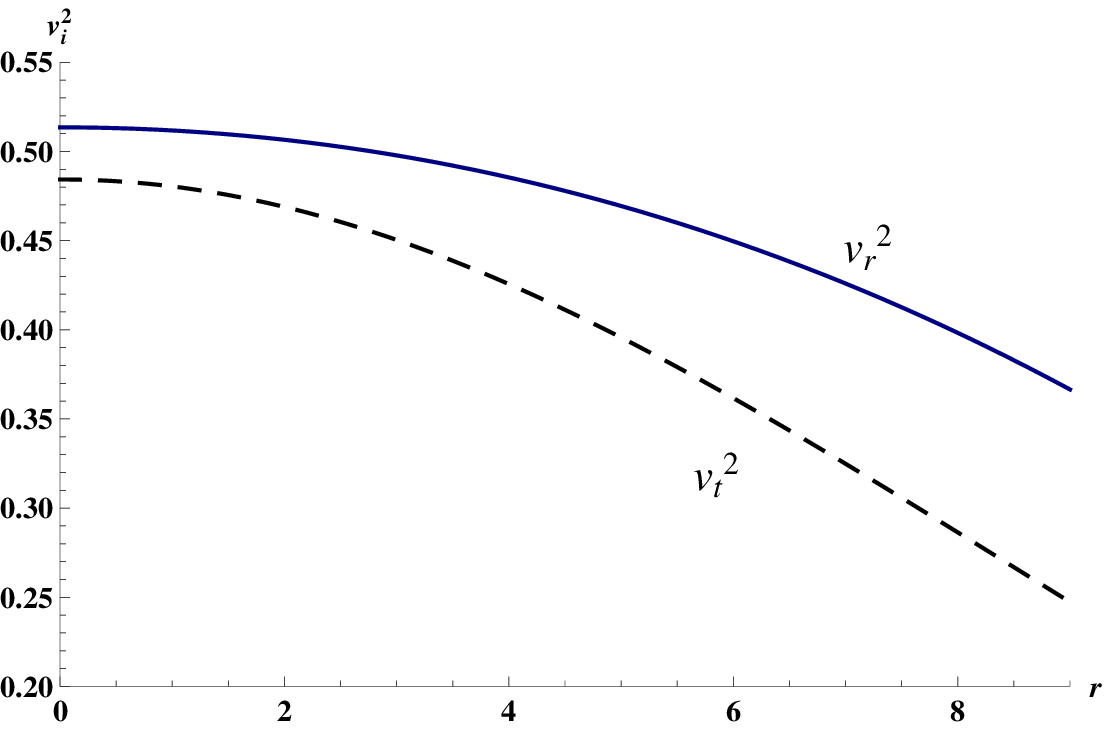}
\figcaption{\label{spe}   Variation of square of sound speeds with radius.}
\end{center}

\begin{center}
\includegraphics[width=5cm]{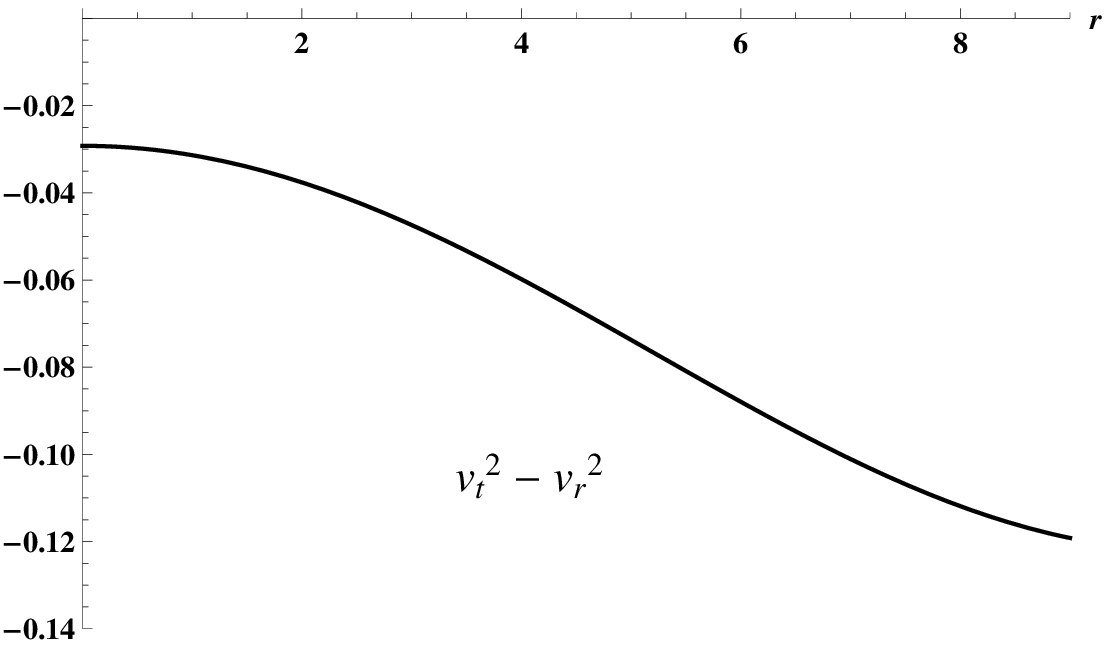}
\figcaption{\label{stab}   Variation of stability factor with radius.}
\end{center}

\begin{center}
\includegraphics[width=5cm]{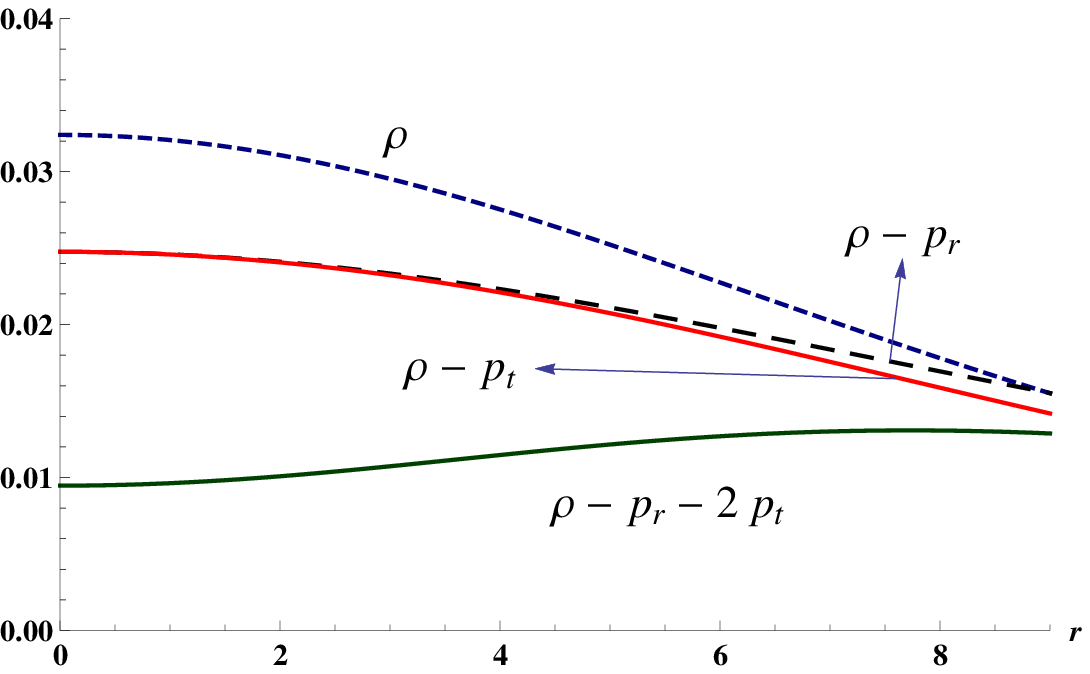}
\figcaption{\label{ener}   Variation of energy conditions with radius.}
\end{center}

\begin{center}
\includegraphics[width=5cm]{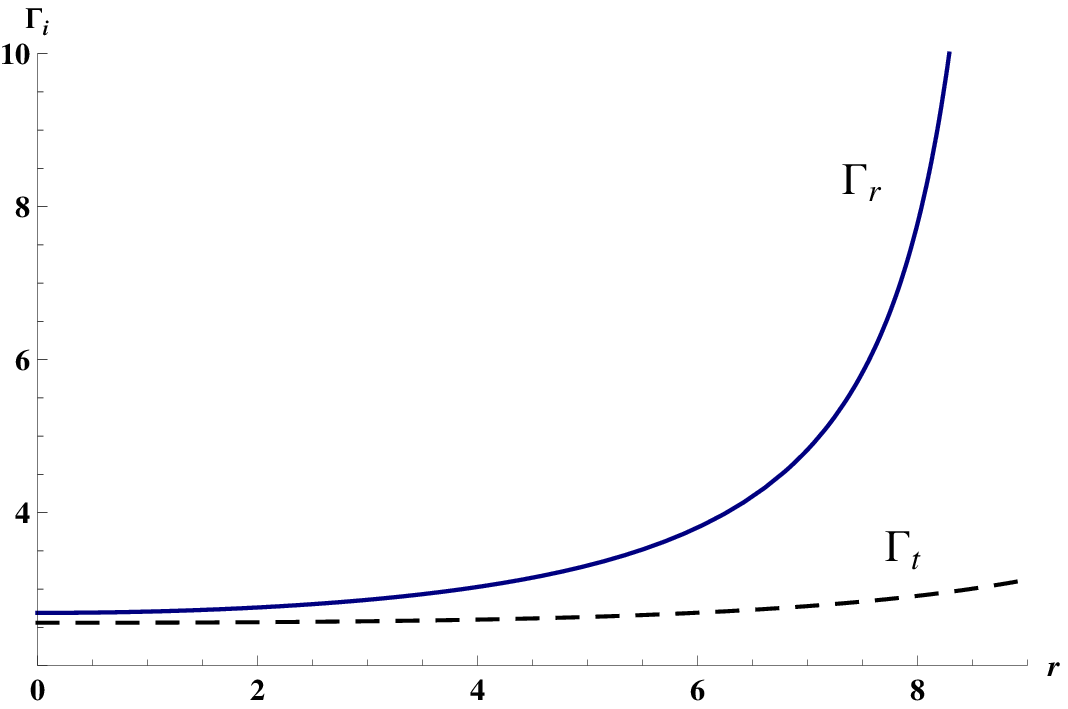}
\figcaption{\label{gam}   Variation of adiabatic index with radius.}
\end{center}

\end{multicols}

\newpage
\begin{center}
\tabcaption{ \label{tab1}  Optimization of masses, radii, Buchdahl limit, surface redshift and compare with observed values.}
\footnotesize
\begin{tabular*}{170mm}{@{\extracolsep{\fill}}cccccccccc}
\toprule 
\multicolumn{1}{c}{Stars} & $a$ (km$^{-2}$) & \multicolumn{1}{c}{$r_b$ (km)} & \multicolumn{1}{c}{$M/M_\odot$} & \multicolumn{1}{c}{$2M/r_b$} & \multicolumn{1}{c}{$z_s$} & $M_{obs}/M_\odot$ & $R_{obs}$ (km) & \multicolumn{1}{c}{Type} & Ref. \\
\hline
Cen X-3   & 0.00147\hphantom{0}  & 8.7\hphantom{0} & 1.21 & 0.278\hphantom{0} & 0.177 & 1.21 $\pm$ 0.21 & -- &  NS & \cite{ash} \\
PSR J0348+0432 & 0.000751 & 13\hphantom{00} & 2.01 & 0.309\hphantom{0} & 0.203 & 2.01 & 13$\pm$2 & NS &\cite{anto} \\
PSR B0943+10 & 0.00077\hphantom{0}  & 2.6\hphantom{0} & 0.02 & 0.0155 & 0.008  & 0.02 & 2.6 &  QS  & \cite{yue}\\
XTE J1739-217  & 0.000934 & 10.9 & 1.51 & 0.277\hphantom{0} & 0.176 & 1.51 & 10.9 & QS & \cite{zhang} \\
\bottomrule
\end{tabular*}%
\end{center}

\begin{multicols}{2}

\begin{center}
\includegraphics[width=5cm]{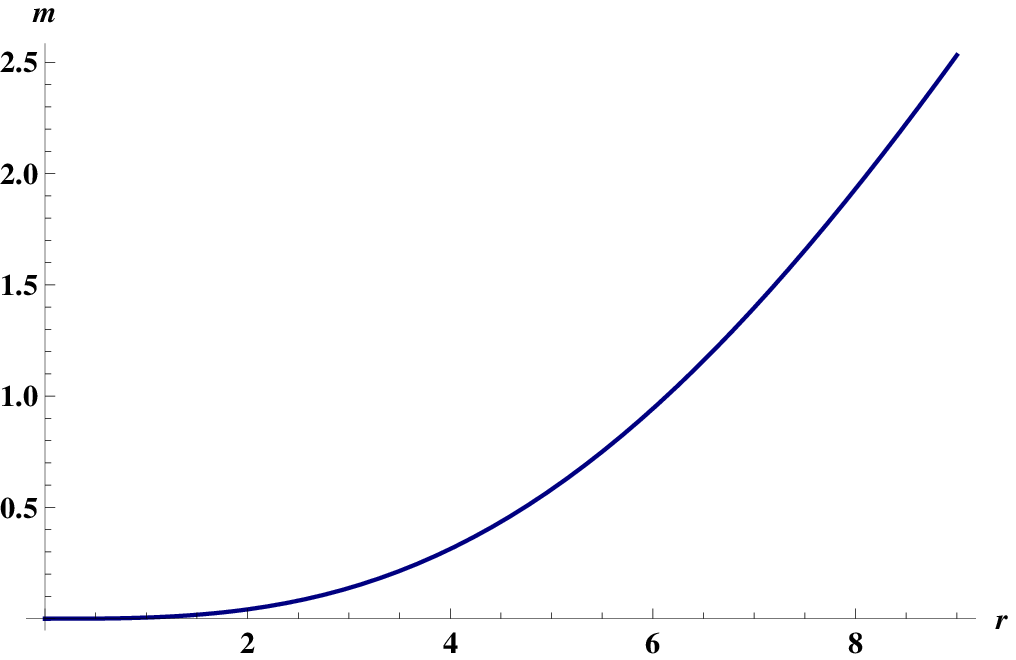}
\figcaption{\label{maf}   Variation of mass with radius.}
\end{center}

\begin{center}
\includegraphics[width=5cm]{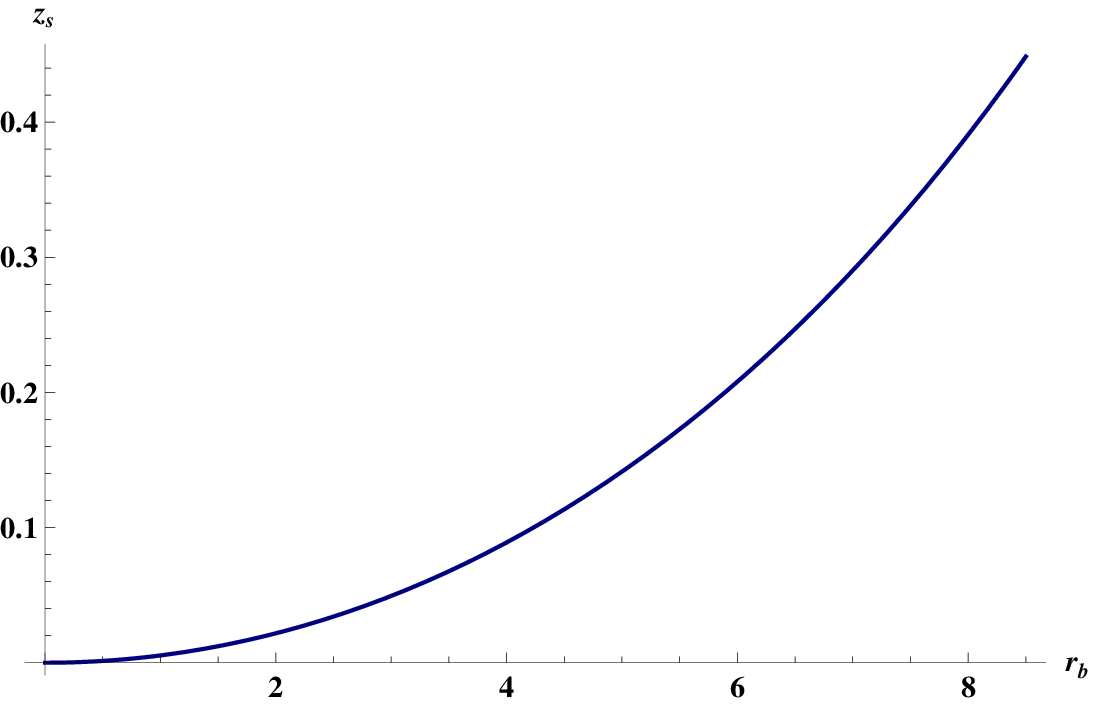}
\figcaption{\label{reds}   Variation of surface redshift with radius.}
\end{center}

\begin{center}
\includegraphics[width=5cm]{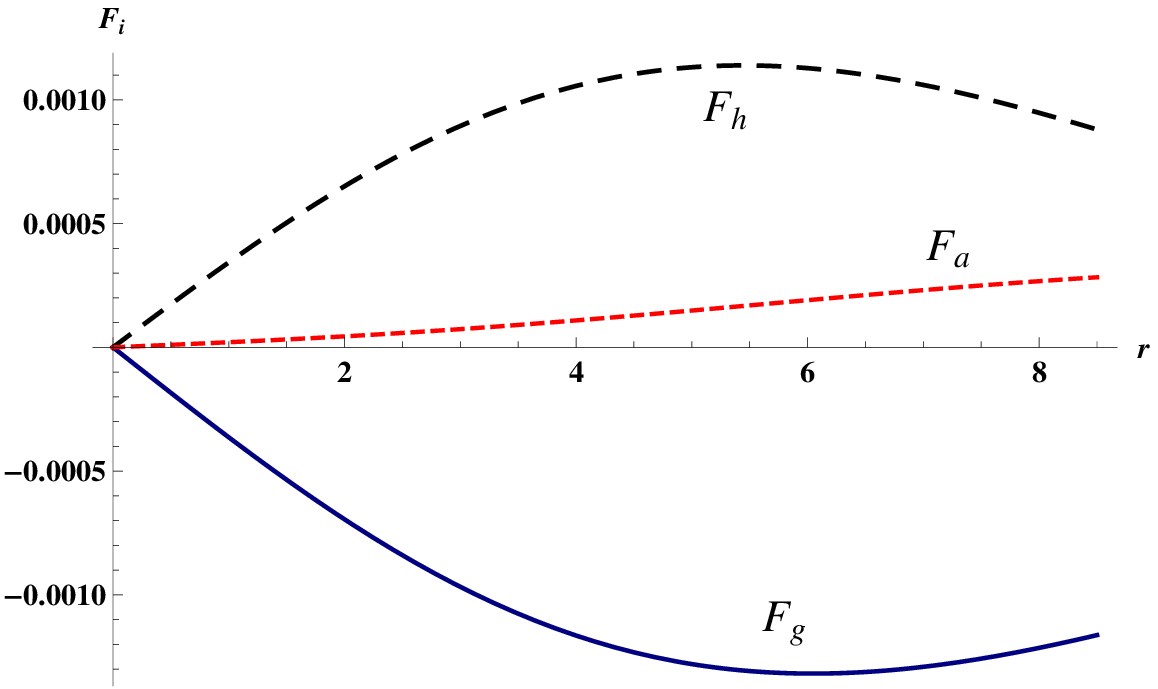}
\figcaption{\label{tov1}   Counter-balancing of three different forces acting in an anisotropic fluid sphere.}
\end{center}

Using the boundary condition (\ref{bou1}-\ref{bou3}),  we get
\begin{eqnarray}
A &=& {2-ar_b^2 \over 4(1+ar_b^2)}- {B \over \sqrt{a}}\bigg[3\sqrt{2}~\tanh^{-1}\left(\sqrt{{4+ar_b^2 \over 6}}\right)  \nonumber \\
&& -\sqrt{12+3ar_b^2}\bigg]\\
{2M \over r_b} & = & 1-{\big(2-ar_b^2\big)^2 \over 4\big(1+ar_b^2\big)^2}\\
{A \over B} &=& \bigg[3 a r_b^2 \sqrt{a r_b^2+4}~ \sqrt{\frac{a r_b^2 (a r_b^2+4)}{(a r_b^2-2)^2}}-6 \sqrt{a r_b^2+4} \nonumber \\
&& \sqrt{\frac{a r_b^2(a r_b^2+4)}{(a r_b^2-2)^2}}~\bigg]^{-1}  \times \nonumber \\
&&  \bigg[36 \sqrt{2}~ r_b~\tanh ^{-1}\left(\sqrt{{a r_b^2+4 \over 6}}\right)  +9 \sqrt{2}~ a r_b^3 \nonumber \\
&& \tanh^{-1}\left(\sqrt{{a r_b^2+4 \over 6}}\right) -16 \sqrt{3}~ r_b \sqrt{a r_b^2+4} \nonumber \\
&& -\sqrt{3} a r_b^3 \sqrt{a r_b^2+4}\bigg]
\end{eqnarray}

Now the gravitational red-shift at the stellar surface is given by
\begin{eqnarray}
z_s &=& e^{-\nu_b/2}-1 = {2(1+ar_b^2) \over 2-ar_b^2}-1
\end{eqnarray}

For physically stable static configuration, the energy condition like Null Energy Condition (NEC), Weak Energy Condition (WEC), Strong Energy Condition (SEC) and Dominant Energy Condition needs to satisfy throughout the interior region i.e.
\begin{eqnarray}
\rho\ge 0;~\rho-p_r \ge 0;~\rho-p_t \ge 0; \nonumber \\
\rho-p_r-2p_t \ge 0;~ \rho \ge (|p_r|,~|p_t|)
\end{eqnarray}

\section{Results and Conclusions}

It has been observed that the physical parameters $\big(p_r,~p_t,~\rho,$ $~p_r/c^2\rho,~p_t/c^2\rho,$ $~v_r^2,~v_t^2 \big)$ are positive at the center and within the limit of realistic equation of state and monotonically decreasing outward (Fig. \ref{rho1}$-$\ref{perho},  \ref{spe}). However the metric potentials, anisotropy, surface red-shift, mass-function and $\Gamma$ are increasing outward which is necessary for a physically viable configuration (Fig. \ref{met}, \ref{ani}, \ref{gam}-\ref{reds}).

Furthermore, our presented solution satisfies all the energy condition which is needed by a physically possible configuration. The Strong Energy Condition (SEC), Weak Energy Condition (WEC), Null Energy Condition (NEC) and Dominant Energy Condition (DEC) is shown in Fig. \ref{ener}. The stability factor $v_t^2-v_r^2$ must lies in between $-1$ and 0 for stable and 0 to 1 for unstable configuration. Therefore the presented solution satisfies stability condition (Fig. \ref{stab}).

The decreasing nature of pressures and density is further justified by their negativity of their gradients, Fig. \ref{grd}. The solution represents a static and equilibrium configuration as the force acting on the fluid sphere is counter-balancing each other. For an anisotropic stellar fluid in equilibrium the gravitational force, the hydro-static pressure and the anisotropic force are acting through TOV-equation and they are counter-balancing to each other, Fig. \ref{tov1}.

Using this solution, we have presented some models of well-known compact stars and compare there observed masses and radii with our calculated values Table \ref{tab1}. Indeed our presented models are in good agreement with the experimentally observed values. Hence the presented solution might have astrophysical significance in the future.

\vspace{-1mm}
\centerline{\rule{80mm}{0.1pt}}

\end{multicols}

\clearpage
\end{CJK*}
\end{document}